\documentclass[12pt,preprint]{aastex}
 
\shorttitle{Distant Red Galaxies}  
\shortauthors{Conselice et al.} 
 
\def\jk{$(J-K)_{\rm vega} > 2.3$}

\def\solm{M$_{\odot}\,$}

\def\mass{$10^{11}$ M$_{\odot}\,$}      
\def\hmass{$10^{11.5}$ M$_{\odot}\,$}      

\begin{document}      

\title{AEGIS: The Diversity of Bright Near-IR Selected Distant Red Galaxies}

\author{C. J. Conselice$^{1}$,  J. A. Newman$^{2,3}$,  A. Georgakakis$^{4}$, O. Almaini$^{1}$, A. L. Coil$^{3,5}$, M.C. Cooper$^{6}$, P. Eisenhardt$^{7}$, S. Foucaud$^{1}$, A. Koekemoer$^{8}$, J. Lotz$^{9}$, K. Noeske$^{10}$, B. Weiner$^{11}$, C.N.A Willmer$^{5}$}

\altaffiltext{1}{School of Physics and Astronomy, University of Nottingham, NG7 2RD, UK}
\altaffiltext{2}{Lawrence Berkeley National Laboratory, Berkeley, CA 94720}
\altaffiltext{3}{Hubble Fellow}
\altaffiltext{4}{Imperial College, London}
\altaffiltext{5}{Stewart Observatory, University of Arizona, Tucson, AZ}
\altaffiltext{6}{University of California, Berkeley}
\altaffiltext{7}{NASA Jet Propulsion Laboratory, California Institute of Technology}
\altaffiltext{8}{Space Telescope Science Institute, Baltimore, MD}
\altaffiltext{9}{Goldberg Fellow, National Optical Astronomy Observatory, Tucson, AZ 85726}
\altaffiltext{10}{University of California, Santa Cruz, Santa Cruz, CA}
\altaffiltext{11}{Department of Physics, University of Maryland, College Park, MD 20742}

\begin{abstract}

We use deep and wide near infrared (NIR) imaging from the Palomar 
telescope combined with DEEP2 spectroscopy and Hubble Space Telescope 
(HST) and Chandra Space Telescope imaging to investigate the 
nature of galaxies that 
are red in NIR colors.  We locate these `distant red galaxies' (DRGs) 
through
the color cut $(J-K)_{\rm vega}~>~2.3$ over 0.7~deg$^{2}$, where we find 
1010 DRG candidates down to $K_{\rm s}~=~20.5$. We combine 95 high 
quality spectroscopic redshifts with photometric redshifts from BRIJK 
photometry to determine the redshift and stellar mass distributions for 
these systems, and morphological/structural and X-ray 
properties for 107 DRGs in the Extended Groth Strip.  We find that many 
bright $(J-K)_{\rm vega}~>~2.3$ galaxies with 
$K_{\rm s}<$~20.5 are at redshifts $z~<~2$, with 64\% between $1<z<2$. 
The stellar mass distributions for these galaxies is broad, ranging from 
10$^{9}~-~10^{12}$~\solm, but with most $z~>~2$ systems massive 
with M$_{*}~>~10^{11}$~\solm.  HST imaging shows that the structural 
properties and morphologies of DRGs are also diverse, with the 
majority elliptical/compact (57\%), and the remainder
edge-on spirals (7\%), and peculiar galaxies (29\%).  
The DRGs at $z < 1.4$ with high quality
spectroscopic redshifts are generally compact, with small half-light 
radii, and span a range in rest-frame optical properties.  
The spectral energy distributions for these objects differ 
from higher redshift DRGs: they are bluer by one
magnitude in observed $(I-J)$ color.  A pure IR color selection of high 
redshift populations is not sufficient to identify unique populations, and 
other colors, or spectroscopic redshifts are needed to produce homogeneous 
samples.

\end{abstract}

\section{Introduction}

Uncovering the formation mechanisms of the most massive galaxies 
has a long history and has largely driven the field of galaxy formation and
evolution.  Traditional analyses of stellar populations in nearby
 ellipticals reveal that the bulk of the stars in the most
massive galaxies formed early, within a few Gyr of the Big
Bang (e.g., Trager et al. 2000).  Observing distant galaxies is 
complementary to nearby galaxy age-dating, as it
allows us to directly observe galaxies while they are forming.   Early
detections of high redshift galaxies revealed that UV bright star forming
galaxies at $z \sim 3$ are dominated by relatively low stellar mass
systems ($\sim 10^{10}$~\solm) (Papovich et al. 2005).  As
the most massive galaxies today are $>10^{11.5}$~\solm, this implies
that high$-z$ galaxies grow by a factor of 5-10 through some process,
such as merging (e.g., Conselice et al. 2003; Conselice 2006).

It seems likely that at least some progenitors of today's massive galaxies 
are not bright in the UV, and a population of near infrared (NIR) selected
'Distant Red Galaxies' (DRGs) 
was described by Saracco et al. (2001) and Franx et al. (2003), who 
suggested that some of these
systems are the progenitors of the most massive present day galaxies.  
These DRGs are
selected using a single color cut, $(J-K)_{\rm vega} > 2.3$, and thus
require near infrared (NIR) imaging to detect.   While it has been
proposed that DRGs are massive $z > 2$ galaxies, the bright end of this
population has yet to be studied, and thus we do not yet know the 
general characteristics of red galaxies selected in the near infrared.

  We investigate this problem  by examining the redshifts, masses, 
morphologies and spectra of DRGs in the Extended Groth Strip (EGS).  
We study 1010 bright DRGs with $K_{\rm s} < 20.5$ within the EGS,
and two other fields that are part of our NIR survey with the 
Palomar telescope.  We examine the structures and morphologies of 
DRGs that are coincident between the ACS imaging in the EGS
and our NIR survey.  We find that a high fraction of DRGs at 
$K_{\rm s} < 20.5$ 
are at $z < 2$, with the majority between $1 < z < 2$, including 95 (10\%)
with spectroscopic redshifts.  We present an initial study of these 
NIR selected galaxies at $z<1.4$ and conclude that they are small, 
low mass, AGN/star formation dominated galaxies.  Furthermore, we find 
a diversity in the morphological properties of DRGs at all redshifts, 
suggesting that there is a wide diversity of properties for galaxies 
selected with the $(J-K)_{\rm vega} > 2.3$ color cut, perhaps even larger 
than that seen for traditional EROs (Moustakas et al. 2004).
In this paper we assume the following cosmology: 
H$_{0}$~=~70~km~s$^{-1}$~Mpc$^{-1}$, 
$\Omega_{\lambda}$ = 0.7, and $\Omega_{\rm m}$ = 0.3, and use Vega magnitude
units throughout.

\section{Data and Sample}

The data used in this paper originate from multiple data sets partially
associated with the All-wavelength Extended Groth Strip International 
Survey (AEGIS) (Davis et al. 2006). The main source of data 
is from the Palomar Observatory Wide-Field Infrared Survey (POWIR) (Bundy
et al. 2005a,b; Conselice et al. in prep), while
other sources include DEEP2 spectroscopy (Davis et al. 2003), 
one orbit Hubble Space Telescope (HST) imaging using the
Advanced Camera for Surveys (ACS) of the EGS (Lotz et al. 2006), and an X-ray 
survey of the region (Nandra et al. 2006 in prep).  The 
POWIR data were obtained from September 2002 until
October 2005 at the Palomar Observatory 5 meter using the Wide-field Infrared
Camera (WIRC).  For the J and K
data we cover a total of $\sim 0.7$ deg$^{2}$ to 5-$\sigma$
point source sensitivities of $K_{\rm s} = 20.5-21.5$ and $J = 22.5-23$.  
Within the EGS, where our highest quality data exist, we image 
0.2 deg$^{2}$ to a 5 sigma depth of $K_{\rm s} = 21.1$ and $J = 23$ 
for point sources.    To construct a sample
of DRGs which is nearly complete in the K-band, we limit our
study of DRGs to those at K$_{\rm s} < 20.5$.

We select DRGs through the use of a 
$(J-K)_{\rm vega} > 2.3$ cut. To supplement the existing 95 DEEP2 
spectroscopic redshifts, we calculate 
photometric redshifts derived from BRIJK photometry using the ANNz method   
for $z < 1.4$ systems, and hyper-z for those at $z > 1.4$.  
The photometric redshift 
accuracy at $z < 1.4$ for all galaxies is quite good,
with $\delta z$/$z$ = 0.05, based on comparisons to $> 10,000$ 
spectroscopic redshifts. The subset of 95 DRGs with $z < 1.4$ spectroscopy 
have a slightly softer accuracy of $\delta z$/$z$ = 0.07, with very few 
catastrophic mismatches. We test our $z > 1.4$ photometric redshifts by 
comparing to 44 galaxies in the Steidel et al. (2004) BM/BX 
$1.5 < z < 2.5 $ population, where we find a lower accuracy of 
$\delta z$/$z$ = 0.22.

We use the spectral energy distributions (SEDs) and 
spectroscopic/photometric redshifts to calculate stellar masses utilizing
the methods outlined in Bundy et al. (2005a,b).   
Our star formation histories are 
parameterized by a single exponentially declining star 
formation  event. We vary
the metallicity, dust extinction and age of these various models to calculate
the distribution of stellar masses using a Chabrier IMF (see Bundy et al. 2005a,b for a detailed discussion).

\section{Results}

\subsection{Basic Redshift and Mass Characteristics}

We find a total of 1010 DRGs, brighter than $K_{\rm s} = 20.5$, over our 
0.7 deg$^{2}$ 
(2520 arcmin$^{2}$) Palomar survey area, including two fields in addition 
to the EGS.  In comparison to previous work, van Dokkum et al. (2006) studied
$\sim 200$ DRGs over 400 arcmin$^{2}$, FIRES 14 DRGs over
100 arcmin$^{2}$, and Papovich et al. (2006) 153 DRGs over
130 arcmin$^{2}$, all with similar DRG number densities at $K_{\rm s} < 20.5$
as found in our study.   Out of our 1010
NIR selected DRGs, we determine that 55 are stars based
on their optical colors and unresolved structures. The DRG redshift
distribution is shown in Figure~1.  A total of 95 of these DRGs have 
secure spectroscopic 
redshifts taken with the DEIMOS spectrograph as part of the DEEP2 survey 
(Davis et al. 2003).  This gives us a lower limit of $\sim 10$\% on 
$z < 1.4$ galaxies in pure photometrically selected DRG 
samples to $K_{\rm s} = 20.5$.  Using these redshifts with our 
calibrated photometric redshifts at $z < 1.4$, we find that $\sim 70$\% of
$K_{\rm s} < 20.5$ DRGs are at $z < 1.4$, but most of the DRG population 
(64\%) is between $1 < z < 2$.
Objects at $z > 2$ represent $\sim 4$\% of the $K_{\rm s} < 20.5$ 
DRG population, while
$\sim 25\%$ of the population is found between $0.5 < z < 1$.  This
implies that bright DRGs have a broad redshift distribution, which is also
found in other recent studies (Reddy et al. 2005; Papovich et al. 2006; 
Grazian et al. 2006).
 
We plot the stellar mass distribution for our sample in Figure~1b. 
From this we conclude that
$(J-K)_{\rm vega} > 2.3$ selected galaxies do not uniquely sample 
high-mass galaxies, 
with a low-redshift `contamination' rate approximately
similar to what Franx et al. (2003) estimate based on
Smail et al. (2002).  DRGs however do select high
redshift systems, generally at $z > 0.8$, and have
a redshift distribution broader than
the ERO population (Moustakas et al. 2004).  A $(J-K)_{\rm vega} > 2.3$ 
selection at $z \sim 1.5$ measures approximate
$(B-I)$ colors similar to dusty or old galaxies at
those redshifts, based on 6 Gyr single burst stellar population models.
In Figure~2 we plot the $(J-K)$ vs. $K_{\rm s}$ 
diagram for all
DRGs at $z > 1.5$ that have stellar masses $>$ \mass. The most
massive galaxies with masses $>$ \hmass are shown as red boxes
for $1.5 < z < 2.0$ systems, and blue circles for those at $z > 2$.  
The green 
crosses represent galaxies with M$_{*} >$ \mass
at $z > 2$.  The average $(J-K)$ color for all galaxies at $z > 2$ 
with M$_{*} >$ \mass
is $<(J-K)> = 2.3\pm0.62$, while the most massive systems with
M$_{*} >$ \hmass have a mean color of $<(J-K)> = 2.4\pm0.7$.  Although the
average massive galaxy has a similar color as the DRG limit, we find
that a significant number ($\sim50$\%) of all massive galaxies
at $2 < z < 3$ will be missed by the standard DRG selection, in agreement
with van Dokkum et al. (2006).  These results are robust when using 
brighter magnitude cuts where our photometry is more accurate,
and at redder color limits, up to $(J-K)_{\rm vega} > 3.3$. These trends also
remain after we select only galaxies
that are redder than 1~$\sigma$ of the DRG limit, and after
testing our selection through a bootstrap resampling.  That is, the 
diversity in redshifts/masses for bright DRGs is not produced by 
photometric errors bringing up  galaxies into the DRG cut.

We find however that at $z > 2$ the DRG selection limit to
$K_{\rm s}  = 20.5$ does locate very massive galaxies, with an average
stellar mass of $<$M$_{*}$$>$ = $10^{11.7\pm0.2}$ \solm.  The
average DRG at all redshifts has a stellar mass of 
M$_{*} = 10^{10.8\pm0.5}$ \solm, with the spectroscopic sample at 
$z < 1.4$ having an average stellar
mass of M$_{*} = 10^{10.5\pm0.3}$ \solm.
For the remainder of this paper we focus on the DRGs found within the
EGS region, where we have ancillory HST and X-ray data.  

\subsection{Structures and Morphologies of DRGs} 

Due to the ACS overlap with our NIR imaging, we can determine the
morphological properties of 107 DRGs at $K_{\rm s} < 20.5$ up to $z \sim 4$.
We utilize both apparent eye-ball morphologies and quantitative
measurements of size and structure on the F814W ACS imaging in the EGS to
determine what type of galaxies DRGs are.  We classify by eye our galaxies
into the types (with the number of objects in each class): 
ellipticals/compact (61), peculiars (31), disks (0), edge-on disks (7), 
and those too faint to classify (8) following the method described in
Conselice et al. (2005).  The morphologies for the \jk\,
galaxies show a diversity of types, as can be seen in Figure~3, similar
to the situation for lower redshift EROs (Moustakas et al. 2004).  The
most common type are ellipticals/compact making up 57\% of
the total DRG sample, with most of
these (80\%) compact. About 7\% of the sample are composed
of edge-on disks, which are likely red in $(J-K)$ due to dust redenning.
Peculiars account for the remainder (29\%) of the DRG sample. At $z > 2$
there is roughly a similar mixture of ellilptical and peculiars. We have
morphologies for 14 systems with spectroscopic redshifts, 
the majority of which are compact 
(\S 4). We find very little evolution in the morphological distribution
with redshift, with a slight increase in the relative number of
peculiar galaxies at higher redshift.

We also utilized the revised CAS system (Conselice 2003; Conselice et al. 2004)
to quantify the structures of the \jk\, galaxies.  The CAS (concentration,
asymmetry, clumpiness) parameters are a non-parameteric method for
measuring the structures of galaxies as resolved on CCD images 
(Conselice 2003). 
As expected from the eye-ball classifications, we find a diversity of
CAS values.  For the systems with spectroscopic redshifts at $z < 1.4$
we find average values, $<C> = 2.4\pm0.1$, $<A> = 0.26\pm0.11$,
$<S> = 0.09\pm0.06$, which is typical for nearby  normal galaxies, and spirals
with exponential profiles
(Conselice 2003).   We find nearly the same values for galaxies with
$z < 1.4$ photometric redshifts.
For the higher redshift sample at $z > 1.5$ we find that the morphologically
classified ellipticals/compacts are more concentrated with $<C> = 3.0\pm0.5$,
with many systems at $C > 3.3$, similar in morphology to nearby massive 
ellipticals. 
Furthermore, as we are examining $z > 1.4$ systems in the UV, we are likely 
underestimating the rest-frame optical light concentration
(Taylor et al. 2006, submitted).  This is an indication, along with their
high masses, that these $z > 1.5$ systems are likely elliptical progenitors.

There are two caveats to using the F814W band ACS imaging
on these galaxies.  The first is that there are redshift
effects which will change the measured parameters (Conselice et al. 2000a,b; 
Conselice 2003).  Systems at $z > 1.2$ 
are also viewed in the rest-frame ultraviolet, complicating 
comparisons to nearby galaxies viewed in the optical.  There is evidence,
however, that distant galaxies dominated by UV bright star formation look
similar in the rest-frame optical and UV (Windhorst et al. 2002; Papovich 
et al. 2005; Conselice et al. 2005).

\section{Low Redshift $z < 1.4$ DRGs}

We have 95 confirmed high quality spectroscopic redshifts for DRGs
at $z < 1.4$.  About 7\% of these systems are traditional EROs 
with $(R-K)_{\rm vega} > 5$, while $\sim$ 33\% have red optical/NIR colors, 
with $(R-K)_{\rm vega} > 4$. The average magnitude of our spectroscopically 
confirmed lower 
redshift DRGs at $z < 1.4$ is $<$M$_{\rm B}> = -20.48\pm1.27$, with an average 
color of $<(U-B)> = -0.16\pm0.23$, and these systems span the
color-magnitude relation, suggesting a heterogeneous origin.  
Our lower redshift DRGs are a mangitude bluer in 
observed $(I-J)$ color compared to the $z > 1.5$ DRGs.   
They are also roughly a magnitude
brighter than the DEEP2 spectroscopic limit, with an average $<$R$>$ = 23.3.
The average half-light radii of the spectroscopically 
selected $(J-K)_{\rm vega} > 2.3$ galaxies is 0.7$\pm$1 kpc, with an 
average stellar 
mass of 10$^{9.8\pm0.46}$ \solm within the ACS region. 
The concentration indices of these systems are moderate, with 
a average of C = 2.8$\pm$0.4, typical for low mass ellipticals, or disks
(Conselice 2003).  The systems at $z < 1.4$ with photometric redshifts 
in the ACS region have very similar masses, sizes and
morphologies as the spectroscopic sample.  Interestingly, none of these
lower redshift DRGs are face-on disks, which is the most common galaxy type at
$z < 1.4$ (Conselice et al. 2005).

To further  investigate the nature of the $z < 1.4$ objects we combine the 
spectra
of all DRGs with spectroscopy to produce a co-added non-flux 
weighted (Figure~4) spectrum.  This spectrum shows that the $z < 1.4$ DRGs 
with emission lines, host star formation, AGN activity, as well as evidence for
post starbursts with ages 1 Gyr, revealed through strong Balmer absorption 
lines.   Features produced
by star formation, including [OII] and Balmer absorption lines,
are seen, as well as higher
excitation lines such as [NeIII] and [OIII].   The average ratio of
([OIII]$\lambda$5007)/(H$\beta$ $\lambda$4861) $\sim 4-4.5$ after
correcting for H$\beta$ absorption. This suggests a mixture of
star formation and AGNs (Seyfert 2s) could be responsible for
the lower redshift sources (Veilleux \& Osterbrock 1987). The preliminary
X-ray Chandra catalog of the EGS reveals that a lower limit of six $z < 2$ 
DRGs have bright X-ray detections to a limit of 
$\sim 10^{-16}$~erg s$^{-1}$~cm$^{-2}$, all with photo-zs, and soft hardness
ratios.  Other sources could be obscured  AGN in moderate L$_{\rm x}$
systems.  We conclude that $(J-K)_{\rm vega} > 2.3$ systems at 
$z < 1.4$ appear to have average masses, a moderate concentration of light, and
are host to star formation and AGN activity, but otherwise are
heterogenous. 
  
\section{Discussion}

With several large area and deep NIR surveys coming online soon, 
such as UKIDSS and VISTA, it is desirable to understand the properties of
pure photometrically selected NIR samples, such as the DRGs with
\jk. Our results suggest that bright DRGs are a mixed population, as other
smaller studies have previously found (Reddy et al. 2005; Papovich et al. 2006;
van Dokkum et al. 2006). The differences between the original DRG, Franx et al.
(2003), study and ours is due to the depth of the surveys, with Franx
et al. being several magnitudes deeper, although most DRGs studied in
detail thus far are at $K_{\rm s} < 20$.  
Because DRGs have unique SEDs, their photometric redshifts are 
potentially harder to measure accurately, and these results, and others
that utilize photometric redshifts, should be viewed as tentative until
significant numbers of spectroscopic redshifts become available.
This is demonstrated by a high absolute lower limit (10\%) contribution to the 
entire DRG population from $z < 1.4$ galaxies, which tend to be small, 
lower mass galaxies with optical AGN signatures.
The morphological mix derived in this paper is however robust, regardless
of the redshift distribution. The DRG population is morphologically 
dominated by compact galaxies, with edge-on spirals and peculiars making 
up 36\% of the population.   In the
future multi-object near-IR spectrographs will be necessary to make
definitive progress in our understanding of non-UV bright galaxies at $z > 2$.

We thank the members of the AEGIS, Palomar and DEEP2 teams, particularly
Kevin Bundy, for their many
invaluable contributions to the surveys that have made this paper possible.  We
thank Chuck Steidel for allowing us to utilize the spectroscopic
redshifts from his BX/BM survey before publication, and Casey Papovich
and Naveen Reddy for enlightening discussions.  This work
was supported by a NSF Astronomy \& Astrophysics Postdoctoral
Fellowship, PPARC and NASA.  ALC is
supported by NASA through Hubble Fellowship grant HF-01182.01-A.

\clearpage

\begin{figure}
\begin{center}
\vspace{-5cm}
\hspace{-0.5cm}
\rotatebox{0}{
\resizebox{\textwidth}{!}{\includegraphics[bb = 25 25 800 800]{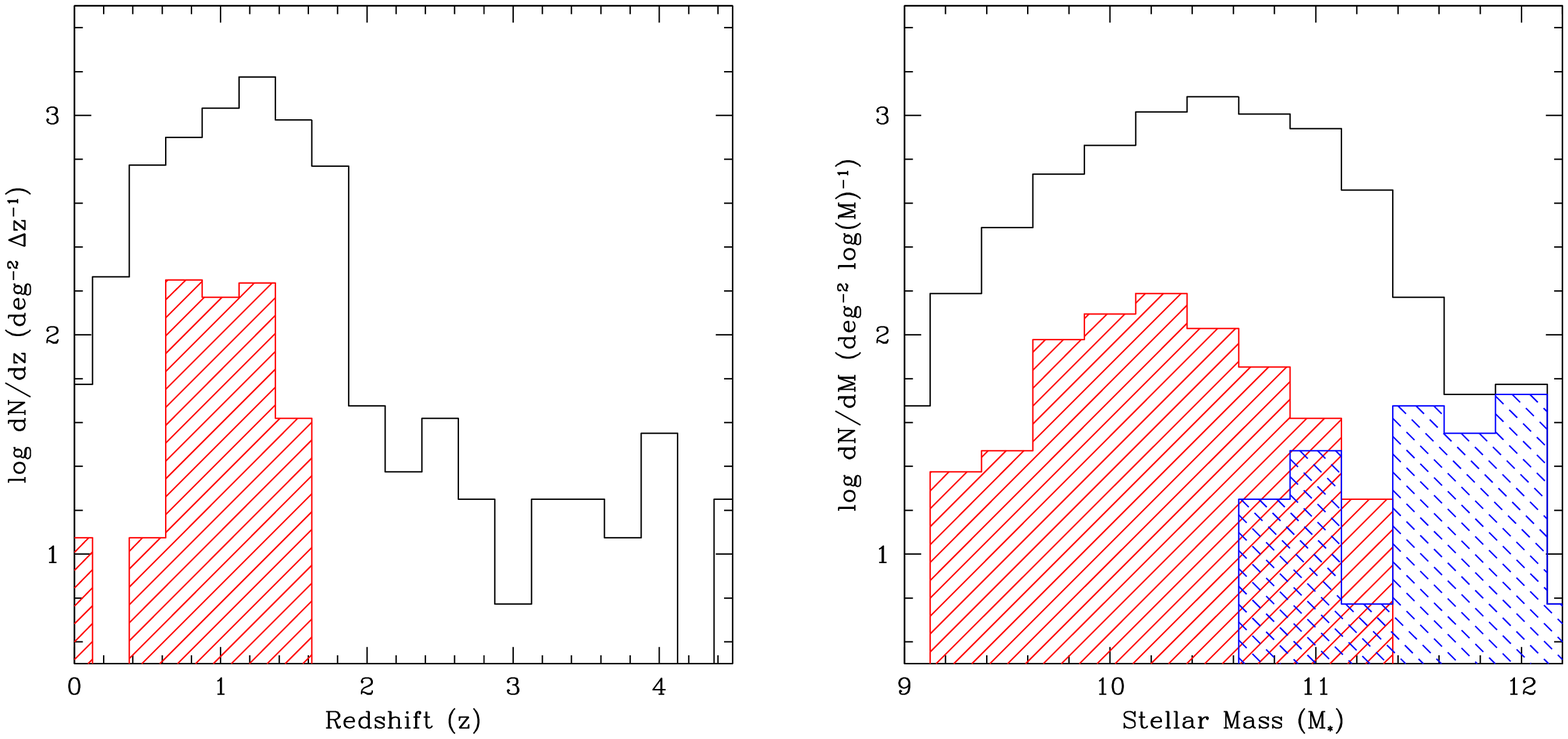}}}
\end{center}
\vspace{-0cm}
\figcaption{Redshift and stellar mass distributions for the 
$K_{\rm s} < 20.5$, DRG $(J-K)_{\rm vega} > 2.3$ population over 
0.7 deg$^{2}$.  The shaded red histogram in the redshift (left) panel is for 
galaxies with reliable spectroscopic redshifts.
The solid shaded red histogram for the mass (right) panel is for galaxies with
spectroscopic redshifts, while the blue dashed shaded histogram shows the
stellar mass distribution for galaxies at $z > 2$.}
\end{figure}   

\clearpage

\begin{figure}
\begin{center}
\vspace{2cm}
\hspace{-0.5cm}
\rotatebox{0}{
\resizebox{\textwidth}{!}{\includegraphics[bb = 25 25 600 600]{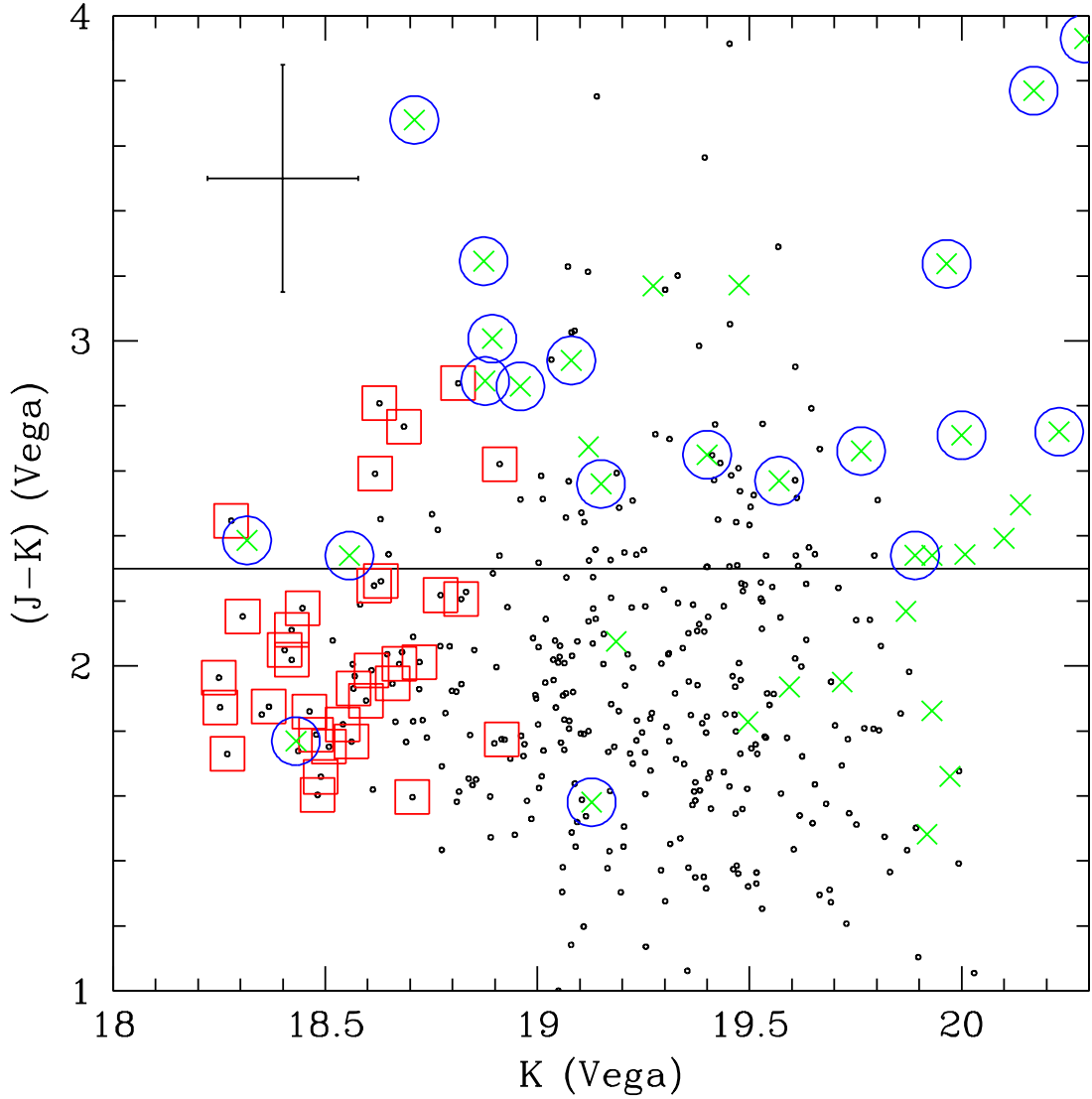}}}
\end{center}
\vspace{-4cm}
\figcaption{The $(J-K)_{\rm vega}$ vs. K$_{\rm s}$
 diagram for objects in the total Palomar
coverage areas.  Every point is for an object at $z > 1.5$ which has
a measured stellar mass with M$_{*} >$ \mass.  Objects surrounded by
red boxes are objects with M$_{*} >$ \hmass at $1.5 < z < 2.0$, while
circled objects are systems with M$_{*} >$ \hmass at $z > 2.0$.  Objects
at $z > 2$ and with M$_{*} >$ \mass are plotted as green crosses.  The
horizontal line is the limit for DRGs. The typical error is shown
in the upper left.}
\end{figure}   

\clearpage

\begin{figure}
\begin{center}
\vspace{-5cm}
\hspace{-0.5cm}
\rotatebox{0}{
\resizebox{\textwidth}{!}{\includegraphics[bb = 25 25 625 625]{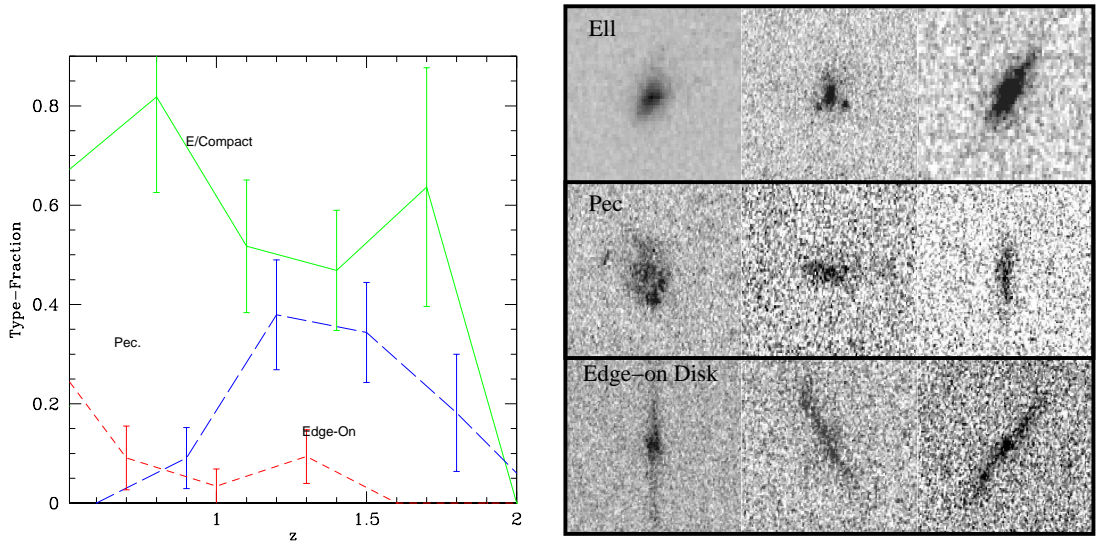}}}
\end{center}
\figcaption{The morphological type fractions for DRGs with
$K_{\rm s} < 20.5$ in the EGS.  The fractions are displayed as: a solid
green line for the compact/ellipticals, 
a long-dashed line for peculiars, and a short-dashed red line for
edge-on disks. Examples of each morphological type are shown on the
right hand side.}
\end{figure}  
 
\clearpage

\begin{figure}
\begin{center}
\vspace{2cm}
\hspace{-0.5cm}
\rotatebox{0}{
\resizebox{\textwidth}{!}{\includegraphics[bb = 25 25 625 625]{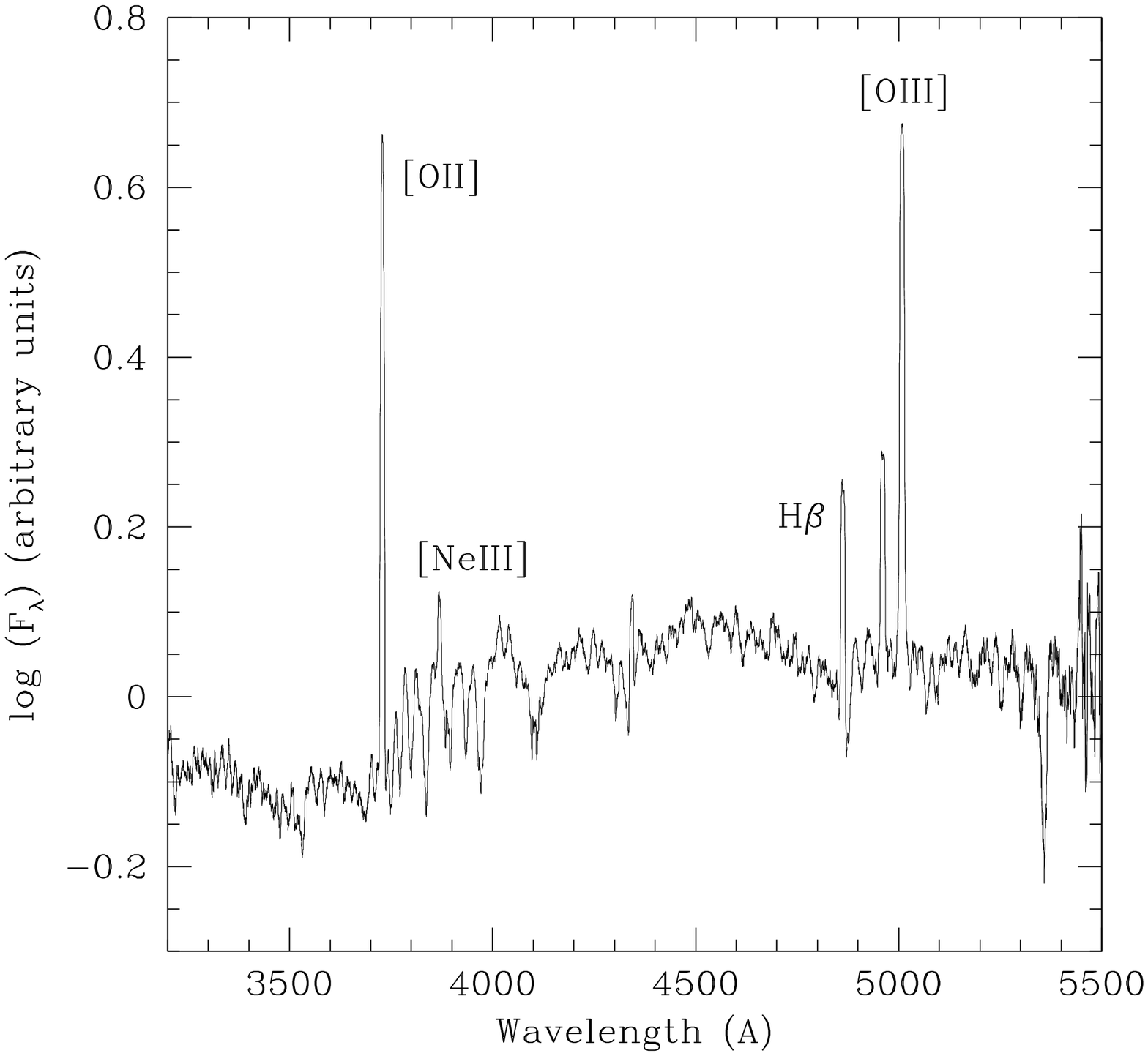}}}
\end{center}
\vspace{-4cm}
\caption{The co-added spectrum for the lower redshift
DRGs with a high quality spectra from the DEEP2 survey.  Lines
due to star formation, such as H$\beta$ can be seen in tandem with
high ionization lines, such as [NeIII], which suggests the presence of
AGN activity.  The ratio of ([OIII]$\lambda$5007)/(H$\beta$ $\lambda$4861)
is also high, which is also indicative of AGN excitation 
for some systems.}
\end{figure} 


\begin{references}

\reference{} Bundy, K., Ellis, R.S., \& Conselice, C.J. 2005a, ApJ, 625, 621
\reference{} Bundy, K., et al. 2005b, astro-ph/0512465
\reference{} Conselice, C.J., Gallagher, J.S., Calzetti, D., Homeier, N., \& Kinney, A. 2000a, AJ, 119, 79
\reference{} Conselice, C.J., Bershady, M.A., \& Jangren, A. 2000b, ApJ, 529, 886
\reference{} Conselice, C.J. 2003, ApJS, 147, 1
\reference{} Conselice, C.J., Bershady, M.A., Dickinson, M., \& Papovich, C. 2003, AJ, 126, 1183
\reference{} Conselice, C.J., et al. 2004, ApJ, 600, 139L
\reference{} Conselice, C.J., Blackburne, J., \& Papovich, C. 2005, ApJ, 620, 564 
\reference{} Conselice, C.J. 2006, ApJ, 638, 686
\reference{} Davis, M., et al. 2003, SPIE, 4834, 161
\reference{} Franx, M. et al. 2003, ApJ, 587, 79L
\reference{} Grazian et al. 2006, astro-ph/0603095
\reference{} Lotz, J., et al. 2006, astro-ph/0602088
\reference{} Moustakas, L.A., et al. 2004, ApJ, 600, 131L
\reference{} Papovich, C., Dickinson, M., Giavalisco, M., Conselice, C.J., \& Ferguson, H.C. 2005, ApJ, 631, 101
\reference{} Papovich, C. et al. 2006, ApJ, 640, 92
\reference{} Reddy, N.A., Erb, D.K., Steidel, C.C., Shapley, A.E., Adelberger, K., \& Pettini, M. 2005, ApJ, 633, 748 
\reference{} Saracco, P. et al. 2001, A\&A, 375, 1
\reference{} Smail, I., Owen, F.N., Morrison, G.E., Keel, W.C., Ivison, R.J., \& Ledlow, M.J. 2002, ApJ, 581, 844
\reference{} Trager, S.C., Faber, S.M., Worthey, G., \& Gonzalez, J.J. 2000, AJ, 119, 1645
\reference{} van Dokkum, P.G., et al. 2006, ApJ, 
\reference{} Veilleux, S., \& Osterbrock, D.E. 1987, ApJS, 63, 295
\reference{} Windhorst, R., et al. 2002, ApJS, 143, 113
\end{references}
\end{document}